\documentclass[12pt,a4paper]{article}
\RequirePackage{amsthm,amsmath,amsfonts,amssymb}
\usepackage[round]{natbib}
\RequirePackage[colorlinks,citecolor=blue,urlcolor=blue]{hyperref}
\usepackage{tikz}
\usepackage{chngcntr}
\usepackage{apptools}
\usepackage[ruled,vlined,linesnumbered]{algorithm2e}
\usepackage{relsize}
\usepackage{array,multirow,makecell}
\usepackage{hhline}
\usepackage{bm}
\usepackage{placeins}
\usepackage{rotating}
\usepackage{caption,subcaption}
\usepackage{dsfont} 
\theoremstyle{plain}
\newtheorem{thm}{Theorem}[section]

\theoremstyle{remark}

\def \m {\mathbf}

\def \beg{\begin{eqnarray}}
\def \en{\end{eqnarray}}
\def \be*{\begin{eqnarray*}}
\def \e*{\end{eqnarray*}}
\def \di{\displaystyle}
\def \bit{\begin{itemize}}
\def \eit{\end{itemize}}
\def \E{\mathbb E}
\def \N{\mathbb N}

\makeatletter
\newcommand{\RemoveAlgoNumber}{\renewcommand{\fnum@algocf}{\AlCapSty{\AlCapFnt\algorithmcfname}}}
\newcommand{\RevertAlgoNumber}{\algocf@resetfnum}
\makeatother
\def \w{\widehat}
\def \t{\tilde}

\def\argmax{\mathop{\mbox{\sl\em argmax}}}

\def\E{\mathbb{E}}

\AtEndDocument{%
  \par
  \medskip
  \begin{tabular}{@{}l@{}}%
    \textsc{Denys POMMERET} (\MakeLowercase{correponding author})\\ 
    \textit{Aix-Marseille University}\\
   \textit{ CNRS, Centrale Marseille, I2M} \\
    \textit{Campus de Luminy} \\
    \textit{13288 Marseille cedex 9}\\
    \textit{Marseille,France}\\
    \textit{E-Mail:} \href{blue}{denys.pommeret@univ-amu.fr}
  \end{tabular}
  
  \vspace{1cm}
    \begin{tabular}{@{}l@{}}%
    \textsc{Yves Ismaël NGOUNOU BAKAM}\\ 
    \textit{Aix-Marseille University}\\
   \textit{ CNRS, Centrale Marseille, I2M} \\
    \textit{Campus de Luminy} \\
    \textit{Marseille,France}\\
    \textit{E-Mail:} \href{blue}{yves-ismael.ngounou-bakam@univ-amu.fr}
  \end{tabular}
  }

\usepackage{sectsty}

\sectionfont{\centering\normalfont\MakeUppercase}

\hypersetup{
	unicode=false,          
	pdftoolbar=true,        
	pdfmenubar=true,        
	pdffitwindow=false,     
	pdfstartview={FitH},    
	pdftitle={My title},    
	pdfauthor={Author},     
	pdfsubject={Subject},   
	pdfcreator={Creator},   
	pdfproducer={Producer}, 
	pdfkeywords={keywords}, 
	pdfnewwindow=true,      
	colorlinks=true,	    
	linkcolor=blue,        
	citecolor=blue,        
	filecolor=blue,        
	urlcolor=blue          
}

\title{Non-parametric Clustering of Multivariate Populations  with Arbitrary Sizes}

\usepackage{authblk}
\date{}
\author[]{BY \vspace{0.3cm}\\Yves I. Ngounou Bakam}
\author[]{Denys Pommeret}
\affil[]{}

\begin{document}

\maketitle
%
\begin{abstract}
We propose a clustering  procedure  to group $K$ populations into  subgroups with the same dependence structure.
The method is adapted to   paired population and  can be  used with panel data. 
It relies on the  
differences between  orthogonal projection coefficients of the $K$ density copulas estimated from the $K$ populations. 
Each cluster is then constituted by populations having  significantly similar dependence structures.
A  recent  test statistic from \cite{Ngounou2} is used  to construct automatically such clusters. 
The procedure is data driven and  depends on the asymptotic level of the test. 
We
illustrate our clustering algorithm via numerical studies
and through two real datasets:  a panel of financial datasets and insurance dataset of losses and allocated loss adjustment expense.

\end{abstract}

 \section*{Keywords} 

Copula coefficients,
data-driven,
Legendre  polynomials,
nonparametric clustering,
smooth test.

 
\section{Introduction and motivations}

The knowledge of the companies that dominate the
capitalization of international stock markets and their
classification can allow portfolio managers a much more
active strategy and a better diversification of risks.
In particular, the knowledge of their dependence structure makes it possible to group together various portfolios with similar risks.

In this paper, we propose a data-driven strategy to regroup portfolios or risks having the same dependence structure.
%
%
%
Their similarities are measured through their  copulas and our procedure is based on  simultaneous multiple comparisons. 
The implementation of our clustering procedure therefore requires a multiple comparison method that has been introduced in \cite{Ngounou}. 
This $K$-sample test   
is a data-driven procedure with a chi-square limit distribution  making the method very fast and very easy to implement. 
The algorithm is based on this test procedure and is also data-driven, depending only  on the asymptotic level of the test. 
The basic idea of this algorithm is to use the test statistics to measure the proximity between populations. If the statistics are close, it is proposed to form a cluster with their associated populations and the test procedure accepts or rejects the validity of the cluster. 

Our method applies to $K (\geq 2)$ iid sample  observed on $K$ populations, eventually paired. This is the case in the considered problem of portfolios. 
Our approach differs from  recent based copulas clustering algorithms as for instance: the clustering methods which rely on hierarchical Kendall copula with Archimedean clusters (see \cite{su2019modelling}, \cite{joe2016multivariate}, among others); a clustering algorithm based on the likelihood of the copula, called CoClust, which has been introduced in \cite{CoClust-article} and further developed and implemented in \cite{CoClustpackage, di2018coclust, di2019clustering}; the clustering algorithms where an iid sample from a finite mixture model is usually considered (see for instance \cite{kosmidis2016model,zhang2019mixtures} and reference therein); the approach which relies on time-varying copula-based estimators via minimization of the value-at-risk (see \cite{de2017dynamic}) and the 
copula-based fuzzy clustering algorithm for spatial time series, called COFUST (see \citet{disegna2017copula}). All these previous works concern parametric copulas and  classify each individual and not populations.

A numerical study first shows the 
good behaviour of the proposed  clustering algorithm.  We then  apply the procedure on financial assets [ a detailler un peu le ou les jeux de donnees ici]. 

The paper is organized as follows: in Section \ref{4.2} we set up notation and we recall the main result of the the test statistic presented in \cite{Ngounou2}, making the paper self-contained.
Section \ref{4.3}  presents  the  clustering algorithm. Section \ref{4.4}  is devoted to the numerical study and Section \ref{4.5}  contains two real-life illustrations.

\section{Notation and test procedure}
\label{4.2}
Let briefly recall here the test procedure proposed in \cite{Ngounou2}.  
Let $\mathbf X=(X_1,\cdots, X_p)$  be a $p$-dimensional  continuous random variable with  joint probability distribution  function (pdf)  $F_{\mathbf X}$ that can be expressed in terms of  copula  as
\beg
\label{cop_chap3}
F_{\m X}(x_1,\cdots, x_p)  & = &
C(F_{1}(x_1),\cdots, F_{p}(x_p) ),
\en
where $F_{j}$  denotes the marginal pdf of $X_j$, and $C$ denotes the  copula associated to $\m X$.
Writing
\be*
U_j= F_{j}(X_j),&&{\rm  for \ }j=1,\cdots, p,
\e*
we have  for all $u_j \in (0,1)$
\be*
C(u_1,\cdots,u_p) & = & F_{\m U}(u_1,\cdots,u_p),
\e*
with $\m U=(U_1,\cdots, U_p)$, and deriving this expression $p$ times with respect to $u_1, \cdots, u_p$, we get an expression of the density copula
\beg
\label{density}
c(u_1,\cdots, u_p) & = & f_{\m U}(u_1,\cdots, u_p),
\en
where $f_{\m U}$ denotes the joint density 
of the vector $\mathbf U$.
Write ${\cal L}=\{L_n ; n\in\N\}$ the set of orthogonal Legendre polynomials (see Appendix \ref{appendlegendre} for more detail). 
Write $\m j=(j_1,\cdots, j_p)\in \N^p$ and define  
\beg
\label{rho_chap3}
\rho_{j_1,\cdots ,j_p} & = &
\E(L_{j_1}(U_1)\cdots L_{j_p}(U_p)),
\en
the $\m j$-th {\it copula coefficient} associated to $\m U$. 
Note that $\rho_{\m 0} =1$ 
where  $\m 0=(0,\cdots, 0)$, and  
 $\rho_{\m j}=0 $ if only one element of $\m j$ is non null. 

The sequence $(\rho_{\m j})_{\m j  \in \N^p_*}$ permits to summarize the copula and we propose a clustering procedure based on the distances between these coefficients. 


In this way assume that we observe $K$  iid samples, possibly paired,
with associated copulas denoted by $C_1, \cdots, C_K$. 

Our aim is to regroup populations having the same copula coefficients, that is, satisfying the following equality
\beg
\label{hypothesis2_chap3}
 H_0: \rho^{(i_1)}_{\m j} = & \cdots &
= \rho^{(i_k)}_{\m j} , \ \ \forall \m j \in \N^p_* ,
\en
where $i_1,\cdots, i_k$ are the label of the tested populations and $\rho^{(i_k)}$ stands for the copula coefficients associated to $C_{i_k}$. 
Clearly if $C_1=\cdots = C_K$  then $H_0$ is immediately satisfied. 
In order to implement our clustering algorithm
we 
propose to use the  statistic based on the estimation of these quantities  proposed in \cite{Ngounou2}. 

We denote by 
$$\mathbf X^{(1)} = (X^{(1)}_{1},\cdots, X^{(1)}_{p}), \cdots, \mathbf X^{(K)} = (X^{(K)}_{1},\cdots, X^{(K)}_{p}),$$
the $K$ continuous random variables associated to the $K$ populations,  with joint cumulative distribution function (cdf) $\mathbf F^{(1)},\cdots, \mathbf F^{(K)}$, and with associated copulas  $C_1, \cdots, C_K$, respectively.
Assume that we observe $K$  iid samples from $\m X^{(1)}, \cdots, \m X^{(K)}$, possibly  paired,  denoted by
$$
(X^{(1)}_{i,1},\cdots, X^{(1)}_{i,p})_{i=1,\cdots, n_1}, \cdots,
( X^{(K)}_{i,1},\cdots, X^{(K)}_{i,p})_{i=1,\cdots, n_K}.
$$
We assume that
\beg
\label{ratiorate}
{\rm for  \ all \  }1\leq k < \ell   \leq K ,&& n_k/(n_k+n_{\ell}) \rightarrow a_{k \ell }, {\rm \ with \ } 0< a_{k \ell } < \infty.
\en
We will denote by $F^{(k)}_{ j}$ the marginal cdf of the  $j$th component of $\m X^{(k)}$ and we write
\begin{align*}
   U^{(k)}_{i,j} & =
   F^{(k)}_{ j}(X^{(k)}_{i,j}).
\end{align*}

For testing (\ref{hypothesis2_chap3}) we first estimate the copula coefficients by
\beg
\label{estimate5}
\w \rho^{(k)}_{j_1\cdots j_p} & = & \di\frac{1}{n_{k}} \di\sum_{i=1}^{n_{k}} L_{j_1}(\w U^{(k)}_{i,1})\cdots L_{j_p}(\w  U^{(k)}_{i,p})),
\en
where
\begin{align*}
 \w  U^{(k)}_{i,j} & =
   \w F^{(k)}_{ j}(X^{(k)}_{i,j}),
\end{align*}
and where $\w F$ denotes the empirical distribution functions associated to $F$.

Considering the null hypothesis $H_0$ as expressed in  (\ref{hypothesis2_chap3}), the  test procedure is based on the   sequences
of  differences
\be*
r^{(\ell,m )}_{\m j} :=  \w \rho^{(\ell)}_{\m j}-\w { \rho}^{(m)}_{\m j},  &&{\rm for \ } 1\leq  \ell \leq m \leq K, {\rm \ and \ } \m j \in \N^p_*,
\e*
with the convention that  $r^{( \ell, m)}_{\m j}=0$ when only one element of $\m j$ is different of zero. 

In order to select automatically the  number of copula coefficients, 
for any vector $\m j=(j_1,\cdots, j_p)$ we denote by
$$\|\m j\|_1=|j_1|+\cdots + |j_p|,$$
its $L^1$ norm and for any integer $d>1$  we write
\be*
{\cal S}(d) & = &\{ \m j\in\N^p; \|\m j\|_1=d {\rm  \ and \ there \ exists \ } k\neq k' {\rm \ such \ that \ }
j_k > 0 {\rm \ and \ }j_{k'} > 0\}.
\e*
The set ${\cal S}(d)$ contains all non null positive  integers $\m j=(j_1,\cdots,j_p)$ with norm $d$  and such that $j_k <d$ for  all $k=1,\cdots, p$.

We also introduce the following set of indexes:
\be*
{\cal V}(K) & = &
\{(\ell,m)\in \N^2 ; 1\leq \ell<m \leq K\}.
\e*
Clearly ${\cal V}(K) $ contains   $v(K)=K(K-1)/2$ elements which represent all the pairs  of populations that we want to compare.

We construct an embedded series of statistics as follows
\be*
V_1 = 
V^{(1,2)}_{D(n)}, \ \  
V_2  =  V^{(1,2)}_{D(n)}+V^{(1,3)}_{D(n)}, \ \
\cdots, \  \
V_{v(K)}  = V^{(1,2)}_{D(n)} + \cdots + V^{(K-1,K)}_{D(n)},
\e*
or equivalently,
\begin{eqnarray*}
V_{k}
& = &
\di\sum_{(\ell,m) \in {\cal V}(K); rank_{\cal V}(\ell,m)\leq k} V^{(\ell,m)}_{D(n)},
 \e*
 where 
\begin{equation} 
\label{sumV_chap3}
V^{(\ell, m)}_k =
 n \di \sum_{\m j \in{\cal  H}(k)}
  (r^{(\ell, m)}_{\m j})^2
\end{equation}
where the set ${\cal  H}(k)$ contains the $k$ first integers of $\N^p$ with respect to the order of ${\cal S}(d)$  
and where 
\begin{equation} \label{rule_chap3}
D(n):= \min\big \{\argmax_{1\leq k \leq d(n)} (  V^{(1, 2)}_{k} - k q_n)  \big\},
\end{equation}
where
$q_n$ and $d(n)$ tend to $ +\infty$ as $n \to +\infty$, $k q_n$ being a penalty term which penalizes the embedded statistics proportionally to the number of  copula coefficients used.

Moreover, we have the following relation: for all $k\geq 1$ and $j=1,\cdots, c(k+1)$
\begin{align*}
V^{(1, 2)}_{c(1)+c(2)+\cdots+c(k)+j}
& =
T^{(1, 2)}_{k+1,j},
\end{align*}
were $c(k)$ denotes the cardinal of ${\cal S=(d)}$ with  the  convention  $c(1)=0$. 

We have $V_1 < \cdots < V_{v(K)}$. The first statistic $V_1$ compares the first two populations 1 and 2. The second statistic $V_2$ compares the populations 1 and 2, and, in addition, the populations 1 and 3. And so on. 
To choose automatically the appropriate number $k$  we introduce the following penalization procedure, mimicking the Schwarz criteria procedure \cite{schwarz2}:

\begin{equation} \label{rule3_chap3}
s(\m n)= \min \Big\{\argmax_{1 \leq k \leq v(K)} \big({V}_{k} - k  p_{n}\big)  \Big\}.
\end{equation}
We make the following assumption:
\begin{description}
\item{{\bf (A)}}  
$d(n)^{(p+4)} = o(q_n)$,
\item{{\bf (A')}}
$d(n)^{(p+4)} = o(p_n)$,
\end{description}
and we recall here main result of \cite{Ngounou2}.
\begin{thm}
\label{thm3_chap3}
Assume that {\bf (A)} and  {\bf (A')}  hold.  Then under $ H_0$,
$s(\m n)$ converges in probability towards 1 as $n \rightarrow +\infty$. 
Moreover, 
$V_{s(\m n)}/\w \sigma^2{(1,2)})$ converges in law towards a $\chi^2_1$ distribution, where  $\w \sigma^2{(1,2)} $ is given in Appendix. 
\end{thm}
It is important to note that if 
$p_n = o(n)$ then the test is consistent against alternative where at least one  copula coefficient differs between two copulas. 

\section{Clustering procedure}
\label{SectionClustering}
\label{4.3}
\subsection{Clustering principle} 
In the sequel we propose to adapt the previous test procedure to obtain a data-driven method  to cluster $K$ populations  into $N$ subgroups characterized  by a common dependence structure. The number $N$ of clusters is unknown and will be automatically chosen by the previous procedure and validated by our testing method.

More precisely, assume that we observe  $K$ iid samples
from $K$ populations, possibly paired.
 The clustering algorithm starts by choosing the two populations that are the most similar in terms of dependence structure, through their copulas.
 In this way, it chooses the smaller two-sample statistic.
If the equality of both associated copulas is accepted these two populations form the first cluster.
Then the algorithm proposes the closer population of this cluster, that is the smaller statistic  having a common population index.
While the test accepts the simultaneous equality of the copulas, the cluster growths.
If the last test is rejected then the cluster is closed and the last rejected population forms a new cluster.
 One can iterate this several times until every sample is associated with a cluster.

\subsection{Clustering algorithm}
We can summarize the clustering algorithm as follows:

\RemoveAlgoNumber
\IncMargin{1.5em}
\begin{algorithm}[H]
\caption{\textbf{K-sample copulas clustering}}
\SetAlgoLined
 \BlankLine
 {\bf Initialization:} 
 $c = 1$. Set $S = \{C_{1},\cdots,C_{K}\}$ and $S_0=\emptyset$ \;
 Select  $\{\ell^\star,m^\star\}={\rm argmin}\{V^{(\ell,m))}_{D(n)}; \ell \neq m \in S \setminus \bigcup_{k=1}^c S_{k-1}\}$ \;
  \BlankLine
 Test $\t H_0$ between all $\rho_{\m j}^{(\ell^\star)}$ and $\rho_{\m j}^{(m^\star)}$ \;
 \eIf{ $\t H_0$ is not rejected}{
  \BlankLine
   $S_1 = \{C_{\ell^\star}, C_{m^\star}\}$\;
   }{
 STOP. There is no cluster. }
 \While{$S\setminus \bigcup_{k=1}^c S_k \neq \emptyset$}{
  \BlankLine
  Select $ \{j^\star\}={\rm argmin}\{T^{(i,j)}_{D(n)}; i\in S_c, j\in S\setminus \bigcup_{k=1}^c S_k\}$\;
  Test $\t H_0$ the simultaneous equality of all the $\rho_{\m j}^{(i)}$, $i\in S_c$ and $\rho_{\m j}^{(j^\star)}$\;
  \BlankLine
  \eIf{ $\t H_0$ not rejected}{
   \BlankLine
    $S_c=S_c\bigcup \{C_{j^\star}\}$\;
   }{
  $S_{c+1} =\{C_{j^\star}\}$\;
  $c=c+1$  \;
  }
 }
\end{algorithm}
\DecMargin{1.5em}
\vspace{0.25cm}
This clustering procedure can solve several complex problems in a very short time and is useful in practice, particularly in risk management and more generally in the world of actuarial science and finance markets by making it possible to detect mutualizable risks and not mutualizable; but also
to build  a well-diversified portfolio.


\subsection{Tuning the algorithm}
\label{tuning}

As evoked in Remark \ref{remark3} we can choose the penalty  $q_n=p_n = \alpha \log(n)$.
We fix $\alpha=1$ in the proofs of this paper for simplicity. 
But in practice  we can empirically improve this tuning  factor by using the following data-driven procedure:

\bit
\item
Assume we observe  $K$ populations.
\item
We merge all populations to get only one (larger) population.
\item
Split randomly this population into $K'>2$ sub-populations.
\item
Clearly   these $K'$ sub-populations  have the same  copula and then  the null hypothesis $\t H_0$ is satisfied.
\item
We then approximate numerically the value of the factor $\alpha>0$ such that the selection rule retains the first component, that is $s(\m n)=1$. 
From Theorem \ref{thm3} this is the asymptotic expected value  under the null.
\item
We can repeat $N$ times such a procedure to get $N$ $K'$-sample under the null.
\eit
Finally we fix
\begin{align*}
\w \alpha & =
\min \{ \alpha>0; \mbox{ such that  } s(\m n) =1   \mbox{ for the previous } N \mbox{ selection rules} \}.
\end{align*}

In our simulation we  fixed arbitrarily $K'=3$, which seems to give a very correct empirical level. Note that this transformation only slightly modified the empirical results.

Concerning the value of $d(n)$, the  condition {\bf (A)} is an asymptotic condition and
from our experience choosing $d(n)=3$ or $4$ is enough to have a very fast procedure which  detects alternatives such that
 copulas differ by a coefficient with a norm less or equal to $d(n)$.

\section{Numerical study of the algorithm}
\label{4.4}

\subsection{Simulation design}

In order to evaluate the performance of the algorithm,
we consider the following classical copulas families: the Gaussian copulas, the Student
copulas, the Gumbel copulas, the Frank copulas,
the Clayton copulas and the Joe copulas which we
denote for hereafter \textit{Gaus}, \textit{Stud},
\textit{Gumb}, \textit{Fran}, \textit{Clay} and
\textit{Joe} respectively.
For the explicit functional forms and properties
of these copulas we refer the reader to
\cite{Nelsen} and \cite{joe1997multivariate}.
For each copula $C$, the sample is generated  with a given
kendall's $\tau$ parameter, and we denote this model briefly by $C(\tau)$.
When $\tau$ is close to zero the variables are close to the independence. Conversely, if $\tau $ is close to $1$ the dependence becomes linear.

\subsection{Clustering simulation}
We consider the following  designs:
\bit
\item
{\bf A100}: $n=100$, $p=3$, $K=6$ populations with 3 groups  $C_{1}=Gumb(0.8)$ and  $C_{2}=C_{3}=Gaus(0.2)$ and $C_{4}=C_{5}=C_{6}=Clay(0.9)$
\item
{\bf A500} = {\bf A100} with $n=500$
\item
{\bf B100}: $n=100$, $p=5$, $K=4$ different populations with
4 groups $C_{1}=Gumb(0.8)$,  $C_{2}=Gaus(0.2)$, $C_{3}=Clay(0.9)$,
$C_{4}=Gumb(1)$
\item
{\bf B500} = B100 with $n=5
00$
\item
{\bf C100}: $n=100$, $p=4$, $K=5$ populations with one group  $C^{(1)}=C^{(2)}=C^{(3)}=C^{(4)}=C^{(5)}=Clay(0.9)$
\item
{\bf C500} = C100 with $n=500$
\item
{\bf D100}: $n=100$, $p=2$, $K=10$ populations with two unbalanced groups
$C_{1}=C_{2}=\cdots = C_{9}=Clay(0.9)$ and
$C_{10}=Gumb(0.9)$
\item
{\bf D500} = D100 with $n=500$
\eit
We applied the clustering algorithm  described in Section  \ref{SectionClustering}. The results are summarized below:
\bit
\item {Results for {\bf A100}:}
\bit
\item
In 82.5 \% of cases the algorithm found 3 groups. In such cases, 74 \% of the time it was the 3 correct groups.
\item
In 11.4 \% of cases the algorithm found 4 groups
\item
In 5 \% of cases the algorithm found 2 groups
\item
In 0.1 \% of cases the algorithm found 5 groups.  \item
Note that the first group (with the Gumbel copula) was well identified  99 \% of the time.
\eit
\item {Results for {\bf A500}:}
The three groups were well identified in 92 \%   of  cases.
In other cases  the algorithm  essentially obtained 4 groups (merging populations of the second and the third group).

\item {Results for {\bf B100}:}
In 78 \% of  cases the null hypothesis was rejected and we obtained 4 different groups.  In other cases the algorithm merged two groups (Clayton with Normal or Clayton with Gumbel) and then proposed 3  clusters.

\item {Results for {\bf C100}:}
In  70 \% of  cases the algorithm found one group.  In other cases it gave two groups.

\item {Results for {\bf D100}:}
More than 80\% of  cases the algorithm found the 2 correct groups.  In other cases the algorithm found 3 group obtained by a rejection of one of the 9 similar populations.
\eit

\begin{table*}[ht]
\begin{tabular}{ccccccc}
\hline
  & 1 & 2 & 3 & 4 & 5 & 6 \\
1 & 100 & 0 & 0 & 0 & 0 & 0 \\
2 &  & 100 & 100 & 0 & 0 & 0 \\
3 & & & 100 & 0 & 0 & 0 \\
4 & & & & 100 & 100 & 100 \\
5 & & & & & 100 & 100 \\
6 & & & & & & 100 \\
\hline
\end{tabular} \hspace*{1cm}
\begin{tabular}{ccccccc}
\hline
  & 1 & 2 & 3 & 4 & 5 & 6 \\
1 & 100 & 0 & 0 & 0 & 0 & 0 \\
2 &  & 100 & 73 & 29 & 30 & 29 \\
3 & & & 100 & 22 & 25 & 21 \\
4 & & & & 100 & 78 & 82 \\
5 & & & & & 100 & 79 \\
6 & & & & & & 100 \\
\hline
\end{tabular}
\caption{ Population associations (in \%)  under model {\bf A100} (n=100). Left: theoretical; Right: observed.  The true associations are $\{1\}; \{2,3\}; \{4,5,6\}$.
}\label{tab_D1}
\end{table*}

\FloatBarrier

\begin{table*}[ht]
\begin{tabular}{ccccccc}
\hline
  & 1 & 2 & 3 & 4 & 5 & 6 \\
1 & 100 & 0 & 0 & 0 & 0 & 0 \\
2 &  & 100 & 100 & 0 & 0 & 0 \\
3 & & & 100 & 0 & 0 & 0 \\
4 & & & & 100 & 100 & 100 \\
5 & & & & & 100 & 100 \\
6 & & & & & & 100 \\
\hline
\end{tabular} \hspace*{1cm}
\begin{tabular}{ccccccc}
\hline
  & 1 & 2 & 3 & 4 & 5 & 6 \\
1 & 100 & 0 & 0 & 0 & 0 & 0 \\
2 &  & 100 & 93 & 0 & 0 & 0 \\
3 & & & 100 & 0 & 0 & 0 \\
4 & & & & 100 & 99 & 99 \\
5 & & & & & 100 & 100 \\
6 & & & & & & 100 \\
\hline
\end{tabular}
\caption{ Population associations (in \%)  under model {\bf A500} (n=500). Left: theoretical; Right: observed. The true associations are $\{1\}; \{2,3\}; \{4,5,6\}$.
}\label{tab_D2}
\end{table*}

\FloatBarrier

\begin{table*}[ht]
\centering
\begin{tabular}{ccccc}
\hline
  & 1 & 2 & 3 & 4  \\
1 & 100 & 0 & 0 & 0  \\
2 &  & 100 & 0 & 0 \\
3 & & & 100 & 0 \\
4 & & & & 100 \\

\hline
\end{tabular} \hspace*{1cm}
\begin{tabular}{ccccc}
\hline
  & 1 & 2 & 3 & 4  \\
1 & 100 & 0 & 0 & 0  \\
2 &  & 100 & 12 & 11  \\
3 & & & 100 & 10  \\
4 & & & & 100  \\
\hline
\end{tabular}

\caption{ Population associations (in \%)  under model {\bf B100}. Left: theoretical; Right: observed. The true associations are $\{1\}; \{2\}; \{3\}; \{4\}$.
}\label{tab_D3}
\end{table*}

\FloatBarrier

\begin{table*}[ht]
\centering
\begin{tabular}{cccccc}
\hline
  & 1 & 2 & 3 & 4 & 5  \\
1 & 100 & 100 & 100 & 100 & 100  \\
2 &  & 100 & 100 & 100 & 100  \\
3 & & & 100 & 100 & 100 \\
4 & & & & 100 & 100  \\
5 & & & & & 100  \\
\hline
\end{tabular} \hspace*{1cm}
\begin{tabular}{cccccc}
\hline
  & 1 & 2 & 3 & 4 & 5  \\
1 & 100 & 97.9 & 98.5 & 98.9 & 99.2  \\
2 &  & 100 & 99.6 & 98.4 & 99.7  \\
3 & & & 100 & 99.4 & 99.1 \\
4 & & & & 100 & 98.7  \\
5 & & & & & 100  \\
\hline
\end{tabular}
\caption{ Population associations (in \%)  under model {\bf C100}. Left: theoretical; Right: observed. The true associations are $\{1,2,3,4,5\}$.
}\label{tab_D4}
\end{table*}

\FloatBarrier

\begin{table*}[ht]
\centering
\begin{tabular}{ccccccccccc}
\hline
  & 1 & 2 & 3 & 4 & 5 & 6 & 7 & 8 & 9 & 10 \\
1 & 100 & 100 & 100 & 100 & 100 & 100 & 100 & 100 & 100 & 0\\
2 &  & 100 & 100 & 100 & 100 & 100 & 100 & 100 & 100 & 0\\
3 & & & 100 &  100  &  100  &  100  &  100 & 100 & 100  & 0 \\
4 & & & & 100 &  100  &  100  & 100  & 100  & 100  & 0\\
5 & & & & & 100 & 100 & 100  &  100 &  100 & 0\\
6 & & & & & & 100 & 100  &  100 &  100 & 0 \\
7 & & & & & & & 100 &  100 &  100 & 0\\
8 & & & & & & & & 100 &  100 &0 \\
9 & & & & & & & & & 100 & 0\\
10 & & & & & & & & & & 100 \\
\hline
\end{tabular} 
\hspace*{1cm}
\begin{tabular}{ccccccccccc}
\hline
  & 1 & 2 & 3 & 4 & 5 & 6 & 7 & 8 & 9 & 10 \\
1 & 100 & 90.2 & 89.8 & 91 & 94.2 & 90.5 & 92 & 97.1 & 89 & 0\\
2 &  & 100 & 94.1 & 92 & 89.9 & 88.7 & 91.3 & 90.9 & 92 & 0\\
3 & & & 100 &  94.4  &  92.2  &  95.6  &  88 & 97.4 & 90  & 0 \\
4 & & & & 100 &  91  &  95.5  & 89.1  & 90  & 93.3  & 0\\
5 & & & & & 100 & 94 & 88.5  &  96 &  97 & 0\\
6 & & & & & & 100 & 89.9  &  91.2 &  88.2 & 0 \\
7 & & & & & & & 100 &  87 &  97.1 & 0\\
8 & & & & & & & & 100 &  96 &0 \\
9 & & & & & & & & & 100 & 0\\
10 & & & & & & & & & & 100 \\
\hline
\end{tabular}
\caption{ Population associations (in \%)  under model {\bf D100}. Up: theoretical; Down: observed. The true associations are $\{1,2, 3, 4, 5, 6, 7, 8, 9\}; \{10\}$.
}\label{tab_D5}
\end{table*}

\FloatBarrier

\section{Real datasets }
\label{4.5}

\subsection{Financial data }

The knowledge of the companies that dominate the
capitalization of international stock markets and their
classification can allow portfolio managers a much more
active strategy and a better diversification of risks.
%
%
We build $33$ portfolios.

From the $500$ component stocks of the $S\&P 500$ stock market
index, which are issued by $500$ large capitalization companies
traded on American stock exchanges, we choose in each sector
following the Global Industry Classification (there exists $11$  stock market sectors)
\begin{itemize}
    \item the stocks index of the $3$ most high weighted companies. We denote $Sih, i=1,\cdots,11$ hereafter,
    \item the stocks index of the  $3$ companies with the middle weight. We denote $Sim, i=1,\cdots,11$ hereafter,
    \item the stocks index of the $3$ companies with the lowest weight. We denote $Si\ell, i=1,\cdots,11$ hereafter.
\end{itemize}
Table \ref{stock-markets_sp500} presents the weight, symbol, company and sector of each selected stock index.

The data employed are weekly closing adjusted prices
from January, $26^{th}$, $2006$ to December, $30^{th}$,
$2021$ for a total of $825$ observations. Data are
available from Yahoo Finance and we consider the rate
returns series by using the standard continuously
compounded return formula. We note that each price of
stock is expressed in the reference country currency.

The application of non-parametric tests of randomness
(\cite{wang2003nonparametric,cho2011generalized,gibbons2014nonparametric})
to these weekly rates of return for each of the $33$
stocks in Table \ref{stock-markets_sp500}
reveals that there is no evidence that these
series are not iid. 
%

We begin by considering the populations (denoted $pop$ in Table \ref{stock-markets_sp500}) of each group (high ($h$), middle ($m$) and lower ($\ell$)). 

Applying the clustering procedure with nominal level $\alpha=5\%$, we obtain $6$,$4$ and $8$ clusters of group $\ell$, group $m$ and group $h$, respectively.
The Figures \ref{group_l_level5}, \ref{group_m_level5} and \ref{group_h_level5} displays the dendrogram of groups (Grp.) $\ell,m$ and $h$ respectively.
In the three dendrogram we observe that the sector Material is isolated. Moreover at $1\%$ level (see Figures \ref{group_h_1level} \ref{group_m_1level} and \ref{group_l_1level}), the number of clusters and the elements of each cluster remain unchanged. But it is clear that moving this level is an interesting way to reduce or increase the number of clusters. 

By looking at the three groups, we now ask whether 
if there are populations in different groups of similar
dependence structure. To this end, we apply the clustering
algorithm to all $33$ populations with $5\%$ nominal level. We get $12$ clusters of populations and the associated dendrogram 
is presented in Figure \ref{sp500_return level5}.
We observe that clusters $C4$,$C5$ and $C9$
contain only the populations of group $\ell$ and
clusters $C8$, $C11$ and $C12$ only the
populations of group $h$. 

We thus obtain a way to group stocks with the same dependence structure into homogeneous portfolios, while forcing these portfolios not to have the same behavior. This allows for risk diversification, for example. 

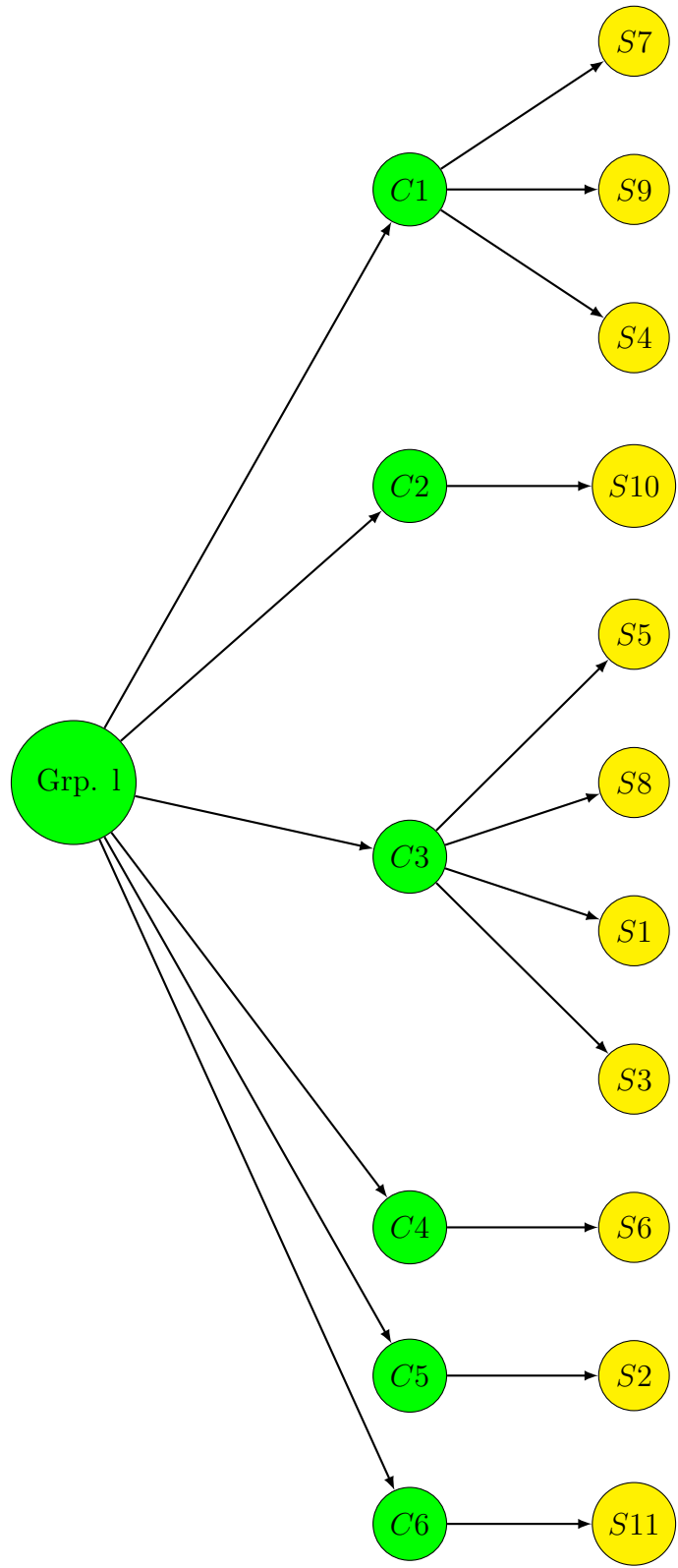
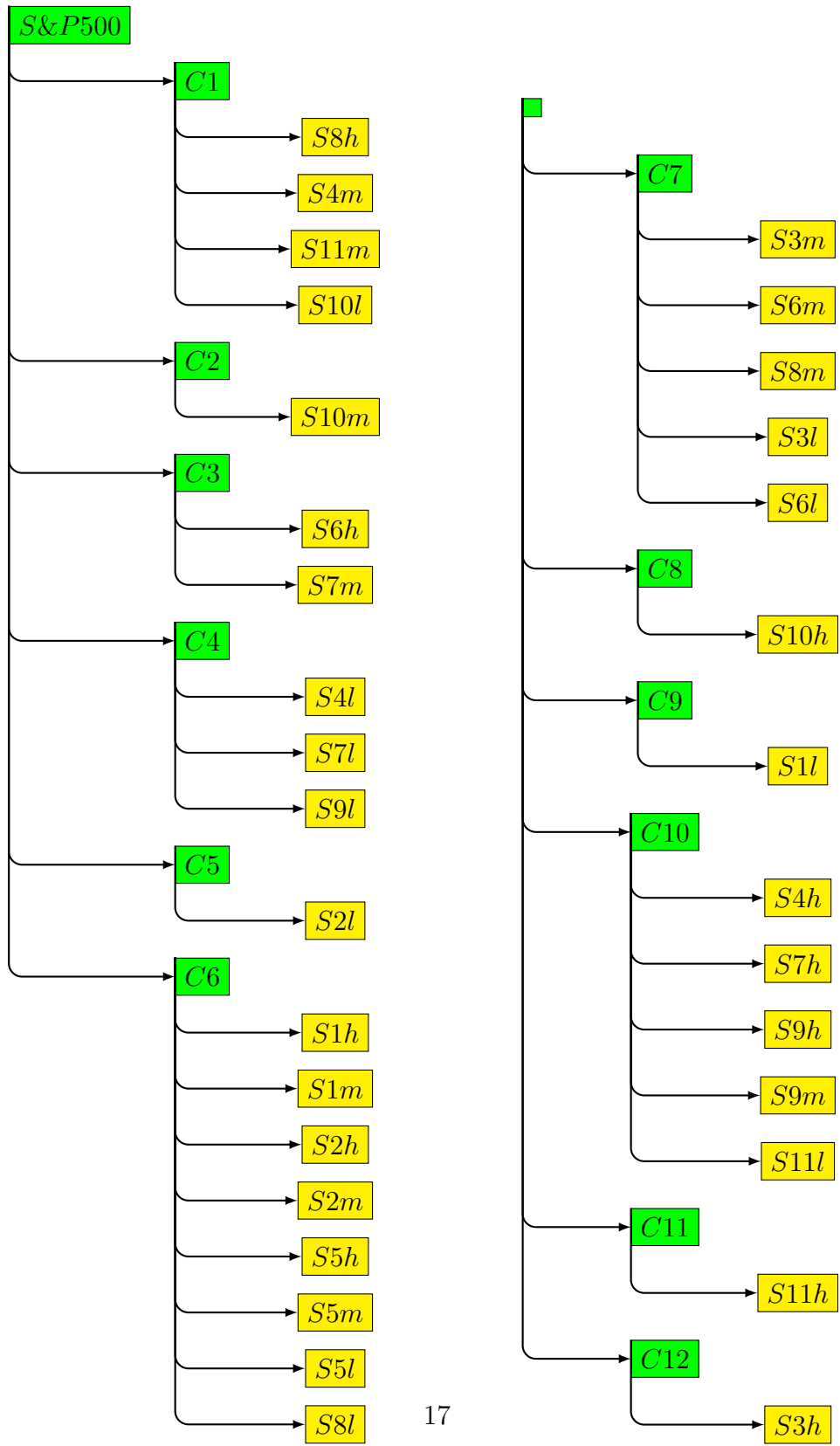
\begin{figure}[!httbp]
\centering

\begin{center}
\begin{tikzpicture}[xscale=1,yscale=1]
\tikzstyle{fleche}=[->,>=latex,thick]
\tikzstyle{noeud}=[fill=green,circle,draw]
\tikzstyle{feuille}=[fill=yellow,circle,draw]
\def\DistanceInterNiveaux{3}
\def\DistanceInterFeuilles{2}
\def\NiveauA{(0)*\DistanceInterNiveaux}
\def\NiveauB{(1.5)*\DistanceInterNiveaux}
\def\NiveauC{(2.5)*\DistanceInterNiveaux}
\def\InterFeuilles{(-1)*\DistanceInterFeuilles}
\node[noeud] (R) at ({\NiveauA},{(5)*\InterFeuilles}) {$\text{ Grp. l}$};
\node[noeud] (Ra) at ({\NiveauB},{(1)*\InterFeuilles}) {$C1$};
\node[feuille] (Raa) at ({\NiveauC},{(0)*\InterFeuilles}) {$S7$};
\node[feuille] (Rab) at ({\NiveauC},{(1)*\InterFeuilles}) {$S9$};
\node[feuille] (Rac) at ({\NiveauC},{(2)*\InterFeuilles}) {$S4$};
\node[noeud] (Rb) at ({\NiveauB},{(3)*\InterFeuilles}) {$C2$};
\node[feuille] (Rba) at ({\NiveauC},{(3)*\InterFeuilles}) {$S10$};
\node[noeud] (Rc) at ({\NiveauB},{(5.5)*\InterFeuilles}) {$C3$};
\node[feuille] (Rca) at ({\NiveauC},{(4)*\InterFeuilles}) {$S5$};
\node[feuille] (Rcb) at ({\NiveauC},{(5)*\InterFeuilles}) {$S8$};
\node[feuille] (Rcc) at ({\NiveauC},{(6)*\InterFeuilles}) {$S1$};
\node[feuille] (Rcd) at ({\NiveauC},{(7)*\InterFeuilles}) {$S3$};
\node[noeud] (Rd) at ({\NiveauB},{(8)*\InterFeuilles}) {$C4$};
\node[feuille] (Rda) at ({\NiveauC},{(8)*\InterFeuilles}) {$S6$};
\node[noeud] (Re) at ({\NiveauB},{(9)*\InterFeuilles}) {$C5$};
\node[feuille] (Rea) at ({\NiveauC},{(9)*\InterFeuilles}) {$S2$};
\node[noeud] (Rf) at ({\NiveauB},{(10)*\InterFeuilles}) {$C6$};
\node[feuille] (Rfa) at ({\NiveauC},{(10)*\InterFeuilles}) {$S11$};
\draw[fleche] (R)--(Ra);
\draw[fleche] (Ra)--(Raa);
\draw[fleche] (Ra)--(Rab);
\draw[fleche] (Ra)--(Rac);
\draw[fleche] (R)--(Rb);
\draw[fleche] (Rb)--(Rba);
\draw[fleche] (R)--(Rc);
\draw[fleche] (Rc)--(Rca);
\draw[fleche] (Rc)--(Rcb);
\draw[fleche] (Rc)--(Rcc);
\draw[fleche] (Rc)--(Rcd);
\draw[fleche] (R)--(Rd);
\draw[fleche] (Rd)--(Rda);
\draw[fleche] (R)--(Re);
\draw[fleche] (Re)--(Rea);
\draw[fleche] (R)--(Rf);
\draw[fleche] (Rf)--(Rfa);
\end{tikzpicture}
\end{center}

\caption[Clustering of group $\ell$ at $5\%$ level. $c_1, \cdots, c_6$ denote the clusters and $s_1, \cdots, s_{12}$ are defined in Table \ref{stock-markets_sp500}.]{Clustering of group $\ell$ at $5\%$ level. $c_1, \cdots, c_6$ denote the clusters and $s_1, \cdots, s_{11}$ are defined in Table \ref{stock-markets_sp500}.}
\label{group_l_level5}
\end{figure}

\begin{figure}[!httbp]
\centering
\begin{center}
\begin{tikzpicture}[xscale=1,yscale=1]
\tikzstyle{fleche}=[->,>=latex,thick]
\tikzstyle{noeud}=[fill=green,circle,draw]
\tikzstyle{feuille}=[fill=yellow,circle,draw]
\def\DistanceInterNiveaux{3}
\def\DistanceInterFeuilles{2}
\def\NiveauA{(0)*\DistanceInterNiveaux}
\def\NiveauB{(1)*\DistanceInterNiveaux}
\def\NiveauC{(2)*\DistanceInterNiveaux}
\def\InterFeuilles{(-1)*\DistanceInterFeuilles}
\node[noeud] (R) at ({\NiveauA},{(5)*\InterFeuilles}) {$\text{ Grp. m}$};
\node[noeud] (Ra) at ({\NiveauB},{(1)*\InterFeuilles}) {$C1$};
\node[feuille] (Raa) at ({\NiveauC},{(0)*\InterFeuilles}) {$S4$};
\node[feuille] (Rab) at ({\NiveauC},{(1)*\InterFeuilles}) {$S11$};
\node[feuille] (Rac) at ({\NiveauC},{(2)*\InterFeuilles}) {$S9$};
\node[noeud] (Rb) at ({\NiveauB},{(4.5)*\InterFeuilles}) {$C2$};
\node[feuille] (Rba) at ({\NiveauC},{(3)*\InterFeuilles}) {$S7$};
\node[feuille] (Rbb) at ({\NiveauC},{(4)*\InterFeuilles}) {$S1$};
\node[feuille] (Rbc) at ({\NiveauC},{(5)*\InterFeuilles}) {$S2$};
\node[feuille] (Rbd) at ({\NiveauC},{(6)*\InterFeuilles}) {$S5$};
\node[noeud] (Rc) at ({\NiveauB},{(8)*\InterFeuilles}) {$C3$};
\node[feuille] (Rca) at ({\NiveauC},{(7)*\InterFeuilles}) {$S8$};
\node[feuille] (Rcb) at ({\NiveauC},{(8)*\InterFeuilles}) {$S6$};
\node[feuille] (Rcc) at ({\NiveauC},{(9)*\InterFeuilles}) {$S3$};
\node[noeud] (Rd) at ({\NiveauB},{(10)*\InterFeuilles}) {$C4$};
\node[feuille] (Rda) at ({\NiveauC},{(10)*\InterFeuilles}) {$S10$};
\draw[fleche] (R)--(Ra);
\draw[fleche] (Ra)--(Raa);
\draw[fleche] (Ra)--(Rab);
\draw[fleche] (Ra)--(Rac);
\draw[fleche] (R)--(Rb);
\draw[fleche] (Rb)--(Rba);
\draw[fleche] (Rb)--(Rbb);
\draw[fleche] (Rb)--(Rbc);
\draw[fleche] (Rb)--(Rbd);
\draw[fleche] (R)--(Rc);
\draw[fleche] (Rc)--(Rca);
\draw[fleche] (Rc)--(Rcb);
\draw[fleche] (Rc)--(Rcc);
\draw[fleche] (R)--(Rd);
\draw[fleche] (Rd)--(Rda);
\end{tikzpicture}
\end{center}

\caption[Clustering of group $m$ at $5\%$ level. $c_1 \cdots c_4$ denote the cluster and $s_1 \cdots s_{11}$ are defined in Table \ref{stock-markets_sp500}.]{Clustering of group $m$ at $5\%$ level. $c_1 \cdots c_4$ denote the cluster and $s_1 \cdots s_{11}$ are defined in Table \ref{stock-markets_sp500}.}
\label{group_m_level5}
\end{figure}

\begin{figure}[!httbp]
\centering

\begin{center}
\begin{tikzpicture}[xscale=1,yscale=1]
\tikzstyle{fleche}=[->,>=latex,thick]
\tikzstyle{noeud}=[fill=green,circle,draw]
\tikzstyle{feuille}=[fill=yellow,circle,draw]
\def\DistanceInterNiveaux{3}
\def\DistanceInterFeuilles{2}
\def\NiveauA{(0)*\DistanceInterNiveaux}
\def\NiveauB{(1)*\DistanceInterNiveaux}
\def\NiveauC{(2)*\DistanceInterNiveaux}
\def\InterFeuilles{(-1)*\DistanceInterFeuilles}
\node[noeud] (R) at ({\NiveauA},{(5)*\InterFeuilles}) {$\text{ Grp. h}$};
\node[noeud] (Ra) at ({\NiveauB},{(0.5)*\InterFeuilles}) {$C1$};
\node[feuille] (Raa) at ({\NiveauC},{(0)*\InterFeuilles}) {$S4$};
\node[feuille] (Rab) at ({\NiveauC},{(1)*\InterFeuilles}) {$S7$};
\node[noeud] (Rb) at ({\NiveauB},{(2.5)*\InterFeuilles}) {$C2$};
\node[feuille] (Rba) at ({\NiveauC},{(2)*\InterFeuilles}) {$S9$};
\node[feuille] (Rbb) at ({\NiveauC},{(3)*\InterFeuilles}) {$S8$};
\node[noeud] (Rc) at ({\NiveauB},{(4)*\InterFeuilles}) {$C3$};
\node[feuille] (Rca) at ({\NiveauC},{(4)*\InterFeuilles}) {$S6$};
\node[noeud] (Rd) at ({\NiveauB},{(5)*\InterFeuilles}) {$C4$};
\node[feuille] (Rda) at ({\NiveauC},{(5)*\InterFeuilles}) {$S11$};
\node[noeud] (Re) at ({\NiveauB},{(6.5)*\InterFeuilles}) {$C5$};
\node[feuille] (Rea) at ({\NiveauC},{(6)*\InterFeuilles}) {$S5$};
\node[feuille] (Reb) at ({\NiveauC},{(7)*\InterFeuilles}) {$S1$};
\node[noeud] (Rf) at ({\NiveauB},{(8)*\InterFeuilles}) {$C6$};
\node[feuille] (Rfa) at ({\NiveauC},{(8)*\InterFeuilles}) {$S2$};
\node[noeud] (Rg) at ({\NiveauB},{(9)*\InterFeuilles}) {$C7$};
\node[feuille] (Rga) at ({\NiveauC},{(9)*\InterFeuilles}) {$S10$};
\node[noeud] (Rh) at ({\NiveauB},{(10)*\InterFeuilles}) {$C8$};
\node[feuille] (Rha) at ({\NiveauC},{(10)*\InterFeuilles}) {$S3$};
\draw[fleche] (R)--(Ra);
\draw[fleche] (Ra)--(Raa);
\draw[fleche] (Ra)--(Rab);
\draw[fleche] (R)--(Rb);
\draw[fleche] (Rb)--(Rba);
\draw[fleche] (Rb)--(Rbb);
\draw[fleche] (R)--(Rc);
\draw[fleche] (Rc)--(Rca);
\draw[fleche] (R)--(Rd);
\draw[fleche] (Rd)--(Rda);
\draw[fleche] (R)--(Re);
\draw[fleche] (Re)--(Rea);
\draw[fleche] (Re)--(Reb);
\draw[fleche] (R)--(Rf);
\draw[fleche] (Rf)--(Rfa);
\draw[fleche] (R)--(Rg);
\draw[fleche] (Rg)--(Rga);
\draw[fleche] (R)--(Rh);
\draw[fleche] (Rh)--(Rha);
\end{tikzpicture}
\end{center}

\caption[]{Clustering of group $h$ at $5\%$ level. $c_1 \cdots c_8$ denote the cluster and $s_1 \cdots s_{11}$ are defined in Table \ref{stock-markets_sp500}}
\label{group_h_level5}
\end{figure}

\bigskip
\begin{figure}[!httbp]
\centering
\noindent
\begin{minipage}[t]{\dimexpr.5\textwidth-.5\columnsep}
\raggedright
\begin{center}
\begin{tikzpicture}[xscale=1,yscale=0.85]
\tikzstyle{fleche}=[->,>=latex,thick,rounded corners=2mm]
\tikzstyle{noeud}=[fill=green,draw]
\tikzstyle{feuille}=[fill=yellow,draw]
\def\DistanceInterNiveaux{2}
\def\DistanceInterFeuilles{1}
\def\NiveauA{(0)*\DistanceInterNiveaux}
\def\NiveauB{(1)*\DistanceInterNiveaux}
\def\NiveauC{(2)*\DistanceInterNiveaux}
\def\InterFeuilles{(-1)*\DistanceInterFeuilles}
\node[noeud] (R) at ({\NiveauA},{(0)*\InterFeuilles}) {$S\&P 500$};
\node[noeud] (Ra) at ({\NiveauB},{(1)*\InterFeuilles}) {$C1$};
\node[feuille] (Raa) at ({\NiveauC},{(2)*\InterFeuilles}) {$S8h$};
\node[feuille] (Rab) at ({\NiveauC},{(3)*\InterFeuilles}) {$S4m$};
\node[feuille] (Rac) at ({\NiveauC},{(4)*\InterFeuilles}) {$S11m$};
\node[feuille] (Rad) at ({\NiveauC},{(5)*\InterFeuilles}) {$S10l$};
\node[noeud] (Rb) at ({\NiveauB},{(6)*\InterFeuilles}) {$C2$};
\node[feuille] (Rba) at ({\NiveauC},{(7)*\InterFeuilles}) {$S10m$};
\node[noeud] (Rc) at ({\NiveauB},{(8)*\InterFeuilles}) {$C3$};
\node[feuille] (Rca) at ({\NiveauC},{(9)*\InterFeuilles}) {$S6h$};
\node[feuille] (Rcb) at ({\NiveauC},{(10)*\InterFeuilles}) {$S7m$};
\node[noeud] (Rd) at ({\NiveauB},{(11)*\InterFeuilles}) {$C4$};
\node[feuille] (Rda) at ({\NiveauC},{(12)*\InterFeuilles}) {$S4l$};
\node[feuille] (Rdb) at ({\NiveauC},{(13)*\InterFeuilles}) {$S7l$};
\node[feuille] (Rdc) at ({\NiveauC},{(14)*\InterFeuilles}) {$S9l$};
\node[noeud] (Re) at ({\NiveauB},{(15)*\InterFeuilles}) {$C5$};
\node[feuille] (Rea) at ({\NiveauC},{(16)*\InterFeuilles}) {$S2l$};
\node[noeud] (Rf) at ({\NiveauB},{(17)*\InterFeuilles}) {$C6$};
\node[feuille] (Rfa) at ({\NiveauC},{(18)*\InterFeuilles}) {$S1h$};
\node[feuille] (Rfb) at ({\NiveauC},{(19)*\InterFeuilles}) {$S1m$};
\node[feuille] (Rfc) at ({\NiveauC},{(20)*\InterFeuilles}) {$S2h$};
\node[feuille] (Rfd) at ({\NiveauC},{(21)*\InterFeuilles}) {$S2m$};
\node[feuille] (Rfe) at ({\NiveauC},{(22)*\InterFeuilles}) {$S5h$};
\node[feuille] (Rff) at ({\NiveauC},{(23)*\InterFeuilles}) {$S5m$};
\node[feuille] (Rfg) at ({\NiveauC},{(24)*\InterFeuilles}) {$S5l$};
\node[feuille] (Rfh) at ({\NiveauC},{(25)*\InterFeuilles}) {$S8l$};
\draw[fleche] (R.north west)|-(Ra);
\draw[fleche] (Ra.north west)|-(Raa);
\draw[fleche] (Ra.north west)|-(Rab);
\draw[fleche] (Ra.north west)|-(Rac);
\draw[fleche] (Ra.north west)|-(Rad);
\draw[fleche] (R.north west)|-(Rb);
\draw[fleche] (Rb.north west)|-(Rba);
\draw[fleche] (R.north west)|-(Rc);
\draw[fleche] (Rc.north west)|-(Rca);
\draw[fleche] (Rc.north west)|-(Rcb);
\draw[fleche] (R.north west)|-(Rd);
\draw[fleche] (Rd.north west)|-(Rda);
\draw[fleche] (Rd.north west)|-(Rdb);
\draw[fleche] (Rd.north west)|-(Rdc);
\draw[fleche] (R.north west)|-(Re);
\draw[fleche] (Re.north west)|-(Rea);
\draw[fleche] (R.north west)|-(Rf);
\draw[fleche] (Rf.north west)|-(Rfa);
\draw[fleche] (Rf.north west)|-(Rfb);
\draw[fleche] (Rf.north west)|-(Rfc);
\draw[fleche] (Rf.north west)|-(Rfd);
\draw[fleche] (Rf.north west)|-(Rfe);
\draw[fleche] (Rf.north west)|-(Rff);
\draw[fleche] (Rf.north west)|-(Rfg);
\draw[fleche] (Rf.north west)|-(Rfh);
\end{tikzpicture}
\end{center}
\end{minipage}
\begin{minipage}[t]{\dimexpr.5\textwidth-.5\columnsep}
\raggedleft
\begin{center}
\begin{tikzpicture}[xscale=1,yscale=1]
\tikzstyle{fleche}=[->,>=latex,thick,rounded corners=2mm]
\tikzstyle{noeud}=[fill=green,draw]
\tikzstyle{feuille}=[fill=yellow,draw]
\def\DistanceInterNiveaux{2}
\def\DistanceInterFeuilles{1}
\def\NiveauA{(0)*\DistanceInterNiveaux}
\def\NiveauB{(1)*\DistanceInterNiveaux}
\def\NiveauC{(2)*\DistanceInterNiveaux}
\def\InterFeuilles{(-1)*\DistanceInterFeuilles}
\node[noeud] (R) at ({\NiveauA},{(0)*\InterFeuilles}) {$$};
\node[noeud] (Ra) at ({\NiveauB},{(1)*\InterFeuilles}) {$C7$};
\node[feuille] (Raa) at ({\NiveauC},{(2)*\InterFeuilles}) {$S3m$};
\node[feuille] (Rab) at ({\NiveauC},{(3)*\InterFeuilles}) {$S6m$};
\node[feuille] (Rac) at ({\NiveauC},{(4)*\InterFeuilles}) {$S8m$};
\node[feuille] (Rad) at ({\NiveauC},{(5)*\InterFeuilles}) {$S3l$};
\node[feuille] (Rae) at ({\NiveauC},{(6)*\InterFeuilles}) {$S6l$};
\node[noeud] (Rb) at ({\NiveauB},{(7)*\InterFeuilles}) {$C8$};
\node[feuille] (Rba) at ({\NiveauC},{(8)*\InterFeuilles}) {$S10h$};
\node[noeud] (Rc) at ({\NiveauB},{(9)*\InterFeuilles}) {$C9$};
\node[feuille] (Rca) at ({\NiveauC},{(10)*\InterFeuilles}) {$S1l$};
\node[noeud] (Rd) at ({\NiveauB},{(11)*\InterFeuilles}) {$C10$};
\node[feuille] (Rda) at ({\NiveauC},{(12)*\InterFeuilles}) {$S4h$};
\node[feuille] (Rdb) at ({\NiveauC},{(13)*\InterFeuilles}) {$S7h$};
\node[feuille] (Rdc) at ({\NiveauC},{(14)*\InterFeuilles}) {$S9h$};
\node[feuille] (Rdd) at ({\NiveauC},{(15)*\InterFeuilles}) {$S9m$};
\node[feuille] (Rde) at ({\NiveauC},{(16)*\InterFeuilles}) {$S11l$};
\node[noeud] (Re) at ({\NiveauB},{(17)*\InterFeuilles}) {$C11$};
\node[feuille] (Rea) at ({\NiveauC},{(18)*\InterFeuilles}) {$S11h$};
\node[noeud] (Rf) at ({\NiveauB},{(19)*\InterFeuilles}) {$C12$};
\node[feuille] (Rfa) at ({\NiveauC},{(20)*\InterFeuilles}) {$S3h$};
\draw[fleche] (R.north west)|-(Ra);
\draw[fleche] (Ra.north west)|-(Raa);
\draw[fleche] (Ra.north west)|-(Rab);
\draw[fleche] (Ra.north west)|-(Rac);
\draw[fleche] (Ra.north west)|-(Rad);
\draw[fleche] (Ra.north west)|-(Rae);
\draw[fleche] (R.north west)|-(Rb);
\draw[fleche] (Rb.north west)|-(Rba);
\draw[fleche] (R.north west)|-(Rc);
\draw[fleche] (Rc.north west)|-(Rca);
\draw[fleche] (R.north west)|-(Rd);
\draw[fleche] (Rd.north west)|-(Rda);
\draw[fleche] (Rd.north west)|-(Rdb);
\draw[fleche] (Rd.north west)|-(Rdc);
\draw[fleche] (Rd.north west)|-(Rdd);
\draw[fleche] (Rd.north west)|-(Rde);
\draw[fleche] (R.north west)|-(Re);
\draw[fleche] (Re.north west)|-(Rea);
\draw[fleche] (R.north west)|-(Rf);
\draw[fleche] (Rf.north west)|-(Rfa);
\end{tikzpicture}
\end{center}

\end{minipage}

\caption[Clustering of $S\&P$ $500$ at $5\%$ level. $c_1 \cdots c_{12}$ denote the cluster and the populations  are defined in Table \ref{stock-markets_sp500}.]{Clustering of $S\&P$ $500$ at $5\%$ level. $c_1 \cdots c_{12}$ denote the cluster and the populations  are defined in Table \ref{stock-markets_sp500}.}
\label{sp500_return level5}
\end{figure}

\FloatBarrier

\begin{table}[htbp]
\centering
\resizebox{17cm}{10cm}{%
\begin{tabular}{|c|c|c|c|c||c|c|c|c|c|}
\hline
Sectors & Pop & Symbols & Companies & Weights &
Sectors & Pop & Symbols & Companies & Weights \\\cline{1-10}
\multirow{9}{*}{\rotatebox{90}{Information Technology}}
&\multirow{3}{*}{\rotatebox{90}{S1h}}
 & AAPL & Apple Inc. & 659.67 & \multirow{9}{*}{\rotatebox{90}{ Consumer Discretionary}}
&\multirow{3}{*}{\rotatebox{90}{S2h}}
 & AMZN &  Amazon.com Inc. & 286.65\\
 & & MSFT & Microsoft Corporation & 582.47 & & & HD & Home Depot Inc. & 91.66\\
 & & NVDA & NVIDIA Corporation & 133.75 & & & MCD & McDonald's Corporation & 53.55\\\cline{2-5}\cline{7-10}
&\multirow{3}{*}{\rotatebox{90}{S1m}}
 & PAYX & Paychex Inc. & 11.26 & &\multirow{3}{*}{\rotatebox{90}{S2m}}
& GPC & Genuine Parts Company & 5.65\\
 & & CDNS & Cadence Design Systems Inc. & 12.26 & & & BBY & Best Buy Co. Inc. & 5.17\\
 & & MCHP & Microchip Technology Incorporated & 11.47 & & & POOL & Pool Corporation & 4.62\\\cline{2-5}\cline{7-10}
&\multirow{3}{*}{\rotatebox{90}{S1l}}
 & FFIV & F5 Inc. & 2.85 & &\multirow{3}{*}{\rotatebox{90}{S2l}}
& PENN &Penn National Gaming Inc. & 1.52\\
 & & JNPR & Juniper Networks Inc. & 2.88 & & & RL & Ralph Lauren Corporation Class A & 1.36\\
 & & DXC & DXC Technology Co. & 2.50 & & & PVH & PVH Corp. & 1.44\\\cline{2-5}\cline{7-10}
\hhline{==========}
\multirow{9}{*}{\rotatebox{90}{ Communication Services}}
&\multirow{3}{*}{\rotatebox{90}{S3h}}
 & GOOGL & Alphabet Inc. Class A & 192.14  & \multirow{9}{*}{\rotatebox{90}{ Financials}}
&\multirow{3}{*}{\rotatebox{90}{S4h}}
 & JPM & JPMorgan Chase \& Co. & 110.32\\
 & & GOOG & Alphabet Inc. Class C & 178.22 & & & BAC & Bank of America Corp & 74.88\\
 & & VZ & Verizon Communications Inc.& 61.40 & & & WFC & Wells Fargo \& Company & 50.74\\\cline{2-5}\cline{7-10}
&\multirow{3}{*}{\rotatebox{90}{S3m}}
 & WBD & Warner Bros. Discovery Inc. Series A & 11.78 & &\multirow{3}{*}{\rotatebox{90}{S4m}}
& MTB & M\&T Bank Corporation & 9.19\\
 & & EA & Electronic Arts Inc. & 11.14 & & & AMP & Ameriprise Financial Inc. & 8.82\\
 & & MTCH & Match Group Inc. & 6.42 & & & TROW & T. Rowe Price Group & 8.47\\\cline{2-5}\cline{7-10}
&\multirow{3}{*}{\rotatebox{90}{S3l}}
 & DISH & DISH Network Corp. Class A & 1.57 & &\multirow{3}{*}{\rotatebox{90}{S4l}}
& ZION & Zions Bancorporation N.A. & 2.46\\
 & & LUMN & Lumen Technologies Inc. & 3.27 & & & BEN & Franklin Resources Inc. & 2.15\\
 & & IPG & Interpublic Group of Companies Inc. & 3.60 & & & IVZ & Invesco Ltd. & 1.89\\\cline{2-5}\cline{7-10}
\hhline{==========}
\multirow{9}{*}{\rotatebox{90}{ Health Care}}
&\multirow{3}{*}{\rotatebox{90}{S5h}}
& UNH & UnitedHealth Group Incorporated & 135.88 &
\multirow{9}{*}{\rotatebox{90}{ Consumer Staples}}
&\multirow{3}{*}{\rotatebox{90}{S6h}}
 & PG & Procter \& Gamble Company & 101.38\\
& & JNJ & Johnson \& Johnson & 135.63 & & & KO & Coca-Cola Company & 71.55\\
 & & PFE & Pfizer Inc. & 86.10 & & & PEP & PepsiCo Inc. & 67.62\\\cline{2-5}\cline{7-10}
&\multirow{3}{*}{\rotatebox{90}{S5m}}
 & BAX & Baxter International Inc. & 10.79 & &\multirow{3}{*}{\rotatebox{90}{S6m}}
& SYY & Sysco Corporation & 12.25\\
 & & A & Agilent Technologies Inc. & 11.20 & & & STZ & Constellation Brands Inc. Class A & 11.49\\
 & & IDXX & IDEXX Laboratories Inc. & 9.61 & & & KR & Kroger Co. & 10.08\\\cline{2-5}\cline{7-10}
&\multirow{3}{*}{\rotatebox{90}{S5l}}
 & UHS & Universal Health Services Inc. Class B & 2.59 & &\multirow{3}{*}{\rotatebox{90}{S6l}}
& HRL & Hormel Foods Corporation & 3.90\\
 & & XRAY & DENTSPLY SIRONA Inc. & 2.46 & & & TAP & Molson Coors Beverage Company Class B & 2.96\\
 & & DVA & DaVita Inc. & 1.71 & & & CPB & Campbell Soup Company &
2.79\\\cline{2-5}\cline{7-10}
\hhline{==========}
\multirow{9}{*}{\rotatebox{90}{Energy}}
&\multirow{3}{*}{\rotatebox{90}{S7h}}
 & XOM & Exxon Mobil Corporation & 117.49 & \multirow{9}{*}{\rotatebox{90}{ Industrials}}
&\multirow{3}{*}{\rotatebox{90}{S8h}}
 & UNP & Union Pacific Corporation & 40.32\\
 & & CVX & Chevron Corporation & 97.79 & & & RTX & Raytheon Technologies Corporation & 41.10\\
 & & COP & ConocoPhillips & 42.47 & & & HON & Honeywell International Inc. & 38.30\\\cline{2-5}\cline{7-10}
&\multirow{3}{*}{\rotatebox{90}{S7m}}
 & OXY & Occidental Petroleum Corporation & 17.82 & &\multirow{3}{*}{\rotatebox{90}{S8m}}
& RSG & Republic Services Inc. & 8.15\\
 & & VLO & Valero Energy Corporation & 15.28 & & & ODFL & Old Dominion Freight Line Inc. & 7.02\\
 & & WMB & Williams Companies Inc. & 12.90 & & & LUV & Southwest Airlines Co. & 7.68\\\cline{2-5}\cline{7-10}
&\multirow{3}{*}{\rotatebox{90}{S7l}}
 & CTRA & Coterra Energy Inc. & 8.18 & &\multirow{3}{*}{\rotatebox{90}{S8l}}
& AOS & A. O. Smith Corporation & 2.31\\
 & & MRO & Marathon Oil Corporation & 6.94 & & & ROL & Rollins Inc. & 2.31\\
 & & APA & APA Corp. & 4.91 & & & ALK & Alaska Air Group Inc. & 1.71\\\cline{2-5}\cline{7-10}
\hhline{==========}
\multirow{9}{*}{\rotatebox{90}{ Utilities}}
&\multirow{3}{*}{\rotatebox{90}{S9h}}
 & NEE & NextEra Energy Inc. & 43.23 & \multirow{9}{*}{\rotatebox{90}{ Materials}}
&\multirow{3}{*}{\rotatebox{90}{S10h}}
 & LIN & Linde plc & 48.08\\
 & & DUK & Duke Energy Corporation & 24.95 & & & SHW & Sherwin-Williams Company & 18.91\\
 & & SO & Southern Company & 22.94 & & & NEM & Newmont Corporation & 15.57\\\cline{2-5}\cline{7-10}
&\multirow{3}{*}{\rotatebox{90}{S9m}}
 & ES & Eversource Energy & 9.09 & &\multirow{3}{*}{\rotatebox{90}{S10m}}
& PPG & PPG Industries Inc. & 8.75\\
 & & DTE & DTE Energy Company & 7.38 & & & ALB & Albemarle Corporation & 8.98\\
 & & EIX & Edison International & 7.54 & & & BALL & Ball Corporation & 6.83\\\cline{2-5}\cline{7-10}
&\multirow{3}{*}{\rotatebox{90}{S9l}}
 & AES & AES Corporation & 4.24 & &\multirow{3}{*}{\rotatebox{90}{S10l}}
& AVY & Avery Dennison Corporation & 4.08\\
 & & NI & NiSource Inc & 3.51 & & & EMN & Eastman Chemical Company & 4.02\\
 & & PNW & Pinnacle West Capital Corporation & 2.53 & & & SEE & Sealed Air Corporation & 2.71\\\cline{2-5}\cline{7-10}
\hhline{==========}
\multirow{9}{*}{\rotatebox{90}{Real Estate}}
&\multirow{3}{*}{\rotatebox{90}{S11h}}
 & AMT & \multicolumn{6}{|c|}{American Tower Corporation} & 33.82 \\
 & & PLD &  \multicolumn{6}{|c|}{Prologis Inc.} & 26.80 \\
 & & CCI &  \multicolumn{6}{|c|}{Crown Castle International Corp} & 23.70\\\cline{2-10}
&\multirow{3}{*}{\rotatebox{90}{S11m}}
 & EQR &  \multicolumn{6}{|c|}{Equity Residential} & 7.55 \\
 & & ARE &  \multicolumn{6}{|c|}{Alexandria Real Estate Equities Inc.} & 6.98\\
 & & VTR &  \multicolumn{6}{|c|}{Ventas Inc.} & 6.51 \\\cline{2-10}
&\multirow{3}{*}{\rotatebox{90}{S11l}}
 & REG &  \multicolumn{6}{|c|}{Regency Centers Corporation} & 2.99\\
 & & FRT &  \multicolumn{6}{|c|}{Federal Realty Investment Trust} & 2.30 \\
 & & VNO &  \multicolumn{6}{|c|}{Vornado Realty Trust} & 1.59\\
\hhline{==========}
\end{tabular}
}
\caption{\label{stock-markets_sp500} $33$ components of S\&P500}
\end{table}

\FloatBarrier

\subsection{Insurance data}

Insurance is an area in which the knowledge of the
dependence structure between several portfolios can be
useful in pricing particularly for risk pooling or price segmentation. 
As an illustration purposes, we consider the well-known
example of pricing insurance contracts involving pairs of
dependent variables which consist to compute the premium
of a reinsurance treaty on a policy with unlimited
liability, some retention level of the losses and a
prorata sharing of ALAEs.
ALAEs in this context are types of insurance company
expenses that are specifically attributable to the
settlement of individual claims such as lawyers' fees and
claims investigation expenses.
The database at issue is the SOA Group Medical Insurance
Large Claims Database over the period $1991–92$ and is
available online at the web page of
\href{https://www.soa.org/resources/experience-studies/20
00-2004/91-92-group-medical-claims/}{Society of
Actuaries}. %
The database includes more than 171,000 claims of $25,000$
or more, representing over $\$10$ billion in total charges
with information collected from $26$ insurers.
Each  row of the database presents a summary of claims for
an individual claimant in fields. Fields include
diagnosis, type of coverage (HMO, PPO, Indemnity, etc.),
claimant status (E-employee or D-dependent), claimant
gender (M-male or F-Female) claimant age and charges split
into hospital and non-hospital.
We refer to \cite{grazier1997group} for a detailed and
thorough description of the data. 
Here, we deal with the $1991$ data of females, insured by
a Preferred Provider Organization (PPO) plan.  We split
the variables losses (hospital charges) and ALAEs (other
charges) by ten-year age groups shown in Table
\ref{soa91}.

\FloatBarrier

\begin{table}[htbp]
\centering
\begin{tabular}{|c|c|c|}
\hline
age groups &  Claimant status & sizes ($n$) \\\hhline{===}
\multirow{2}{*}{$[20,30[$} & D & 426 \\
& E & 568\\\hline
\multirow{2}{*}{$[30,40[$} & D & 967  \\
& E & 1116\\\hline
\multirow{2}{*}{$[40,50[$} & D & 1079  \\
& E & 1177 \\\hline
\multirow{2}{*}{$[50,60[$} & D & 1039 \\
& E & 1136 \\\hline
\multirow{2}{*}{$[60,70[$} & D & 595  \\
& E & 786  \\\hline
\multirow{2}{*}{$[70,80[$} & D & 102  \\
& E & 175 \\\hline
%
\end{tabular}
\caption{\label{soa91} Age groups of females in SOA$91$}
\end{table}

\FloatBarrier

 Applying our algorithm  procedure at $5\%$ level, we
 obtained  four clusters and the dendogram is presented in
 Figure \ref{soa91_F}. 
It appears that the dependence structure of claim charges
change over age where it shows that the status of the
policy holder is irrelevant and that premiums charged to
both types of individuals should be the same if the size of the observations are substantially identical.

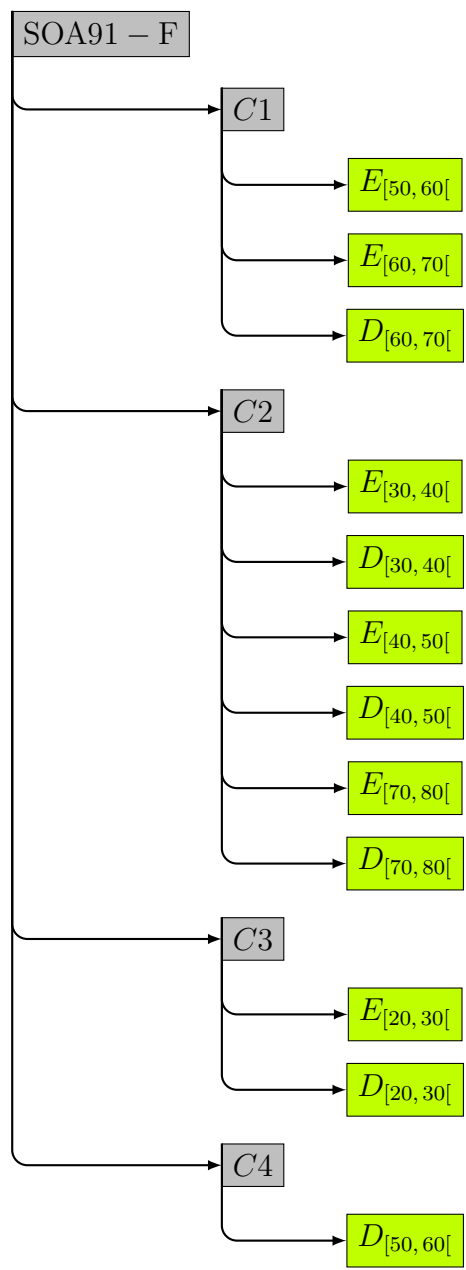
\begin{figure}[!httbp]
\centering

\begin{center}
\begin{tikzpicture}[xscale=1,yscale=1]
\tikzstyle{fleche}=[->,>=latex,thick,rounded corners=2mm]
\tikzstyle{noeud}=[fill=lightgray,draw]
\tikzstyle{feuille}=[fill=lime,draw]
\def\DistanceInterNiveaux{2}
\def\DistanceInterFeuilles{1}
\def\NiveauA{(0)*\DistanceInterNiveaux}
\def\NiveauB{(1)*\DistanceInterNiveaux}
\def\NiveauC{(2)*\DistanceInterNiveaux}
\def\InterFeuilles{(-1)*\DistanceInterFeuilles}
\node[noeud] (R) at ({\NiveauA},{(0)*\InterFeuilles}) {$\text{SOA}91-\text{F}$};
\node[noeud] (Ra) at ({\NiveauB},{(1)*\InterFeuilles}) {$C1$};
\node[feuille] (Raa) at ({\NiveauC},{(2)*\InterFeuilles}) {$E_{[50,\,60[}$};
\node[feuille] (Rab) at ({\NiveauC},{(3)*\InterFeuilles}) {$E_{[60,\,70[}$};
\node[feuille] (Rac) at ({\NiveauC},{(4)*\InterFeuilles}) {$D_{[60,\,70[}$};
\node[noeud] (Rb) at ({\NiveauB},{(5)*\InterFeuilles}) {$C2$};
\node[feuille] (Rba) at ({\NiveauC},{(6)*\InterFeuilles}) {$E_{[30,\,40[}$};
\node[feuille] (Rbb) at ({\NiveauC},{(7)*\InterFeuilles}) {$D_{[30,\,40[}$};
\node[feuille] (Rbc) at ({\NiveauC},{(8)*\InterFeuilles}) {$E_{[40,\,50[}$};
\node[feuille] (Rbd) at ({\NiveauC},{(9)*\InterFeuilles}) {$D_{[40,\,50[}$};
\node[feuille] (Rbe) at ({\NiveauC},{(10)*\InterFeuilles}) {$E_{[70,\,80[}$};
\node[feuille] (Rbf) at ({\NiveauC},{(11)*\InterFeuilles}) {$D_{[70,\,80[}$};
\node[noeud] (Rc) at ({\NiveauB},{(12)*\InterFeuilles}) {$C3$};
\node[feuille] (Rca) at ({\NiveauC},{(13)*\InterFeuilles}) {$E_{[20,\,30[}$};
\node[feuille] (Rcb) at ({\NiveauC},{(14)*\InterFeuilles}) {$D_{[20,\,30[}$};
\node[noeud] (Rd) at ({\NiveauB},{(15)*\InterFeuilles}) {$C4$};
\node[feuille] (Rda) at ({\NiveauC},{(16)*\InterFeuilles}) {$D_{[50,\,60[}$};
\draw[fleche] (R.north west)|-(Ra);
\draw[fleche] (Ra.north west)|-(Raa);
\draw[fleche] (Ra.north west)|-(Rab);
\draw[fleche] (Ra.north west)|-(Rac);
\draw[fleche] (R.north west)|-(Rb);
\draw[fleche] (Rb.north west)|-(Rba);
\draw[fleche] (Rb.north west)|-(Rbb);
\draw[fleche] (Rb.north west)|-(Rbc);
\draw[fleche] (Rb.north west)|-(Rbd);
\draw[fleche] (Rb.north west)|-(Rbe);
\draw[fleche] (Rb.north west)|-(Rbf);
\draw[fleche] (R.north west)|-(Rc);
\draw[fleche] (Rc.north west)|-(Rca);
\draw[fleche] (Rc.north west)|-(Rcb);
\draw[fleche] (R.north west)|-(Rd);
\draw[fleche] (Rd.north west)|-(Rda);
\end{tikzpicture}
\end{center}

\caption{Dendogram of SOA$91$-Female at $5\%$ level. $C_1 \cdots C_{4}$ denote the cluster.}
\label{soa91_F}
\end{figure}

\FloatBarrier

%

%
%

%

\section{Conclusion}

%


\bibliographystyle{plainnat}
\bibliography{clustering}
\vspace{3cm}

\end{document}